# Piezoelectric thin-film super-lattices without using piezoelectric materials

N.D. Sharma[1], C.M. Landis[3], P. Sharma[1,2,*]

[1]Department of Mechanical Engineering, University of Houston, Houston, TX, 77204, U.S.A
[2]Department of Physics, University of Houston, Houston, TX, 77204, U.S.A
[3]Aerospace Engineering and Engineering Mechanics, University of Texas at Austin, TX, 78712, U.S.A.

**Abstract:** In this paper we show that experimentally realizable a*pparently piezoelectric thin-film super-lattices* can be created from non-piezoelectric materials provided an odd-order (e.g. trilayer) stacking sequence is used. The size-dependent mechanism of flexoelectricity, which couples gradients of strain to polarization, allows such a possibility. We present closed-form analytical expressions for the response of various thin-film and super-lattice configurations. We also clarify some of the subtleties that arise in considering interface boundary conditions in the theory of flexoelectricity as well as the relationship of flexoelectricity to the frequently used polarization gradient terms used in modeling ferroelectrics. We find that for certain (optimum) material combinations and length scales, thin film superlattices yielding apparent piezoelectricity close to 75 % of ferroelectric Barium Titanate may be achievable.

## 1. INTRODUCTION AND CENTRAL CONCEPT

In non-centrosymmetric dielectric crystals such as quartz and ZnO, a net electrical dipole moment is generated upon application of uniform strain due to relative displacements between the centers of oppositely charged ions. This well known phenomenon is known as piezoelectricity (Nye, 1985). Formally, the polarization vector is related to the second order strain tensor through the third order piezoelectric tensor as

$$P_i = p_{ijk}\varepsilon_{jk} \qquad (1)$$

Being an odd-ordered tensor, $p_{ijk}$ must vanish for all dielectrics with inversion-center symmetry, thus restricting existence of piezoelectricity to only non-centrosymmetric crystal structures. However, physically, this inversion symmetry of a dielectric unit cell can be broken locally by application of non-uniform strain or the presence of strain gradients. This contribution of macroscopic strain gradient towards induced polarization is known as the flexoelectric effect and can be written as:

$$P_i = \underbrace{p_{ijk}\varepsilon_{jk}}_{=0,\, for\, centrosymmetric\, materials} + \mu_{ijkl}\frac{\partial \varepsilon_{jk}}{\partial x_l} \qquad (2)$$

Here the fourth ordered tensor $\mu_{ijkl}$ is the so-called flexoelectric tensor, and is non-zero for crystals of any symmetry. This implies that under a non-uniform strain, all dielectric materials are capable of producing a polarization. Readers are referred to Tagantsev (1986, 1991) for a review. The microscopic (atomistic) underpinnings of flexoelectricity were recently discussed by one of us (Maranganti and Sharma, 2009) where flexoelectric properties were atomistically calculated for several dielectrics of technological and scientific interest. An interesting example of the flexoelectric response is that of graphene (a manifestly non-piezoelectric material) and clearly elucidated by the atomistic calculations of Dumitrica et. al. (2002) and Kalinin et. al. (2007).

---
[*] Corresponding author: psharma@uh.edu

An estimate for lower bounds of the flexoelectric coefficients was provided by Kogan (1963) to be of the order of $e/a$ ($\approx 10^{-9}$C/m) which was corroborated for the case of an isotropic elastomer by Marvan et. al. (1988). Here $e$ is the electronic charge and $a$ is lattice parameter. Later a simple linear chain model of ions (Marvan and Havranek, 1997) and experiments (Ma and Cross, 2001a) suggested a dependence on the relative permittivity for the case of ordinary dielectrics. For ferroelectric perovskites like PMN, PZT and BST, even in the paraelectric phase, much larger magnitudes ($\approx 10^{-6}$C/m) of flexoelectric coefficients than this lower bound are observed (Ma and Cross, 2001b; 2002; 2003, 2006; Fu et al 2006). Recently, Zubko *et al* (2007) have published the experimental characterization of the complete flexoelectric tensor for $SrTiO_3$.

Several researchers have studied flexoelectricity recently and proposed various applications and consequences of this phenomenon. For example, Catalan et al (2004) have studied the impact of flexoelectricity on the dielectric properties and Curie temperature of ferroelectric materials while Cross and coworkers (1999, 2006) have proposed fabrication of piezoelectric composites without using piezoelectric materials. Eliseev et. al. (2008) have investigated the renormalization in properties of ferroelectric nanostructures due to the spontaneous flexoelectric effect as well as analytical approaches to elucidate size-effects in such nanostructures (Eliseev and Morozovska, 2009). In our previous work (Sharma et al, 2007), we computationally analyzed and demonstrated the possibility of designing such composites through suitable topology, constituent property differences and the selection of optimum feature sizes. Such topologies are hard to realize in practice however. Non-piezoelectric, tapered pyramidal structures on a substrate that "effectively" act as piezoelectric meta-materials have been fabricated in experimental studies by Cross (2006), Fu et al (2006) and Ma and Cross (2006). A strong size-dependent enhancement of the apparent piezoelectric coefficient in materials that are intrinsically piezoelectric has been demonstrated by Majdoub et al (2008a,b) through atomistic calculations. These flexoelectric composites have important technological ramifications such as in actuators, sensors, energy storage and harvesting among others. In a recent work, Majdoub et. al. (2009) demonstrated, through first principles and theoretical calculations, that the so-called dead-layer effect in nanocapacitors may be strongly influenced by flexoelectricity. Several specialized topics have been well-reviewed in a recent book (Yang, 2009).

Page **2** of 25

The central concept behind this paper (a continuation of our previous work, Sharma et. al., 2007) is simple. Consider a composite consisting of two or more different non-piezoelectric dielectric materials. Even under the application of uniform stress, differences in material properties at the interfaces will result in the presence of strain gradients. Those gradients will induce polarization due to the flexoelectric effect. For "properly designed"[1] composites, the net average polarization will be non-zero. Thus, the nanostructure will exhibit an overall electromechanical coupling under uniform stress behaving like a piezoelectric material. The individual constituents must be at the nanoscale since this concept requires very large strain gradients and those (for a given strain) are generated easily only at the nanoscale.

While some general theoretical ideas behind the aforementioned concept were sketched out in a previous work (Sharma et. al., 2007), the homogenization process was crude and the resulting 3D topologies difficult to realize experimentally. In the present work we present closed-form solutions for easily-fabricated thin film super-lattice structures that demonstrate the central concept in a transparent manner. The outline of the paper is as follows: In Section 2, we review the basic theory of flexoelectricity, discuss some subtleties regarding the interfacial boundary conditions and comment on how flexoelectricity relates to the (often used) polarization gradient terms in modeling ferroelectrics. In Section 3, we discuss the symmetry arguments that drive the creation of apparently piezoelectric super-lattices without using piezoelectric materials. In Section 4, we provide general flexoelectricity solutions for the various thin-film layered configurations and calculate the overall electromechanical coupling.

## 2. THEORY OF FLEXOELECTRICITY, RELATION TO POLARIZATION GRADIENT THEORIES AND INTERFACIAL BOUNDARY CONDITIONS

Within the assumptions of the linearized theory for *centrosymmetric* dielectrics, the Helmholtz energy density of deformation and polarization $W^L$ can be assumed to be quadratic function of terms involving small strain $e_{ij}$, polarization $P_i$, polarization gradient $P_{i,j}$ and strain gradient $u_{j,kl}$ (Mindlin, 1972):

$$W^L\left(P_i, e_{ij}, P_{i,j}, u_{j,kl}\right) = \frac{1}{2}a_{kl}P_k P_l + \frac{1}{2}b_{ijkl}P_{i,j}P_{k,l} + \frac{1}{2}c_{ijkl}e_{ij}e_{kl} + d_{ijkl}P_{i,j}e_{kl} \\ + f_{ijkl}P_i u_{j,kl}. \tag{3}$$

Here, $e_{ij}$ are the components of the strain tensor **e** defined as

$$e_{ij} = \frac{1}{2}(u_{i,j} + u_{j,i}), \tag{4}$$

while **a**, **b**, **c**, **d**, **f** are material property tensors. In particular, '**a**' and '**c**' are the familiar second order reciprocal dielectric susceptibility and fourth order elastic constant tensors respectively. The remaining tensors correspond to higher order electro-elastic couplings which do not occur in the classical continuum description of an isotropic elastic dielectric. '**d**', which was introduced by Mindlin (1968) in his theory of polarization gradient, links gradients of polarization to strains while the components of '**f**' are the flexoelectric coefficients.

---
[1] See Section 3

If $\phi$ is the potential of the electric field **E** given by

$$E_i = -\phi_{,i}, \tag{5}$$

then the energy density of **E** must be added to Equation (4) yielding the total potential energy $W$:

$$W = W^L + \frac{1}{2}\varepsilon_0 \varphi_{,i}\varphi_{,j}. \tag{6}$$

Neglecting the effect of charge density as suggested by Askar and Lee (1970), the total electric enthalpy density can be written as

$$\Sigma = W - \left(\varepsilon_0 E_i + P_i\right)E_i, \tag{7}$$

which simplifies to

$$\Sigma = \frac{1}{2}a_{kl}P_k P_l + \frac{1}{2}b_{ijkl}P_{i,j}P_{k,l} + \frac{1}{2}c_{ijkl}e_{ij}e_{kl} + d_{ijkl}P_{i,j}e_{kl} + f_{ijkl}P_i u_{j,kl} \\ -\frac{1}{2}\varepsilon_0 \varphi_{,i}\varphi_{,j} + P_i\varphi_{,i}. \tag{8}$$

The tensor **f** in Equation (8) is related to the tensor **μ** of Equation (2) as (Maranganti and Sharma, 2009)

$$f_{ijkl} = a_{im}\left(\mu_{mjkl} + \mu_{mjlk} - \mu_{mklj}\right). \tag{9}$$

All the tensors corresponding to the material properties are of even order since the restriction to centrosymmetry (i.e., classically non-piezoelectric materials) requires that odd order tensors vanish.

The phenomenon of flexoelectricity in crystalline dielectrics was first predicted by Maskevich and Tolpygo (1957); a phenomenological description was later proposed by Kogan (1963) who included a term coupling the polarization and the strain-gradient in the thermodynamic potential of the form

$$f_{ijkl}P_i u_{j,kl}. \tag{10}$$

Yet another body of work, which parallels the theory of flexoelectricity in some ways, is the polarization gradient theory due to Mindlin (1969, 1971). Based on the long-wavelength limit of the shell-model of lattice dynamics, Mindlin (1969) found that the core-shell and the shell-shell interactions could be incorporated phenomenologically by including the coupling of polarization gradients to strain and the coupling of polarization-gradients to polarization-gradients respectively in the thermodynamic potential (Equations (11a-b))

$$d_{ijkl} P_{i,j} e_{kl},$$
$$b_{ijkl} P_{i,j} P_{k,l}. \tag{11a-b}$$

Material property tensors $\mathbf{d}$ and $\mathbf{b}$ are constants introduced by Mindlin in this polarization gradient theory. The polarization-gradient strain coupling (represented by tensor $\mathbf{d}$) and the polarization strain-gradient coupling (represented by tensor $\mathbf{f}$) are often included in the energy density expression as a Lifshitz invariant (Landau and Lifshitz, 1984) as shown in Equation (12) on account of the fact that total derivatives cannot occur in the expression for energy.

$$h_{ijkl}\left(u_{ij}P_{k,l} - P_k u_{ij,l}\right) \tag{12}$$

This is justified if one considers the following argument. The contribution to the total energy of a finite volume of material including the flexoelectric and the polarization gradient term (only the one involving $\mathbf{d}$) is:

$$\int_V \left(f_{ijkl} P_i u_{j,kl} + d_{ijkl} P_{i,j} u_{k,l}\right) dx. \tag{13}$$

Integration by parts yields:

$$\int_V \left(d_{ijkl} P_{i,j} u_{k,l} - f_{ijkl} P_{i,l} u_{j,k}\right) dx + \text{Boundary terms}. \tag{14}$$

In other words, the governing equations remain unaltered if we use an expression of the form $\left(d_{ijkl} P_{i,j} u_{k,l} - f_{ijkl} P_{i,l} u_{j,k}\right)$ as the energy density. Alternatively in terms of only one of the material tensors (say $\mathbf{h}$),

$$\left(d_{ijkl} - f_{iklj}\right) P_{i,j} u_{k,l}$$
$$= h_{ijkl} P_{i,j} u_{k,l}. \tag{15}$$

The contributions due to the term in the thermodynamic potential involving Mindlin's tensor $\mathbf{d}$ and due to flexoelectricity (involving tensor $\mathbf{f}$) cannot be readily isolated from each other (Maranganti and Sharma, 2009). Thus, mathematically, Mindlin's polarization gradient theory (1968) can be adapted to include the flexoelectric effect (strain gradient-polarization coupling) by replacing the coupling tensor $\mathbf{d}$ by tensor $\mathbf{h}$ as defined in Equation (15). The new tensor $\mathbf{h}$ thus derived represents combination of two fundamentally different coupling phenomena (i) strain-polarization gradient coupling (Mindlin's theory) and (ii) strain gradient – polarization coupling (flexoelectricity).

In order to further elucidate this assertion, we employ the following argument to recover expression (12). Consider

$$\int_V h_{ijkl} P_{i,j} u_{k,l} dx, \qquad (16)$$

which can be decomposed as

$$\int_V \left( \frac{h_{ijkl}}{2} P_{i,j} u_{k,l} + \frac{h_{ijkl}}{2} P_{i,j} u_{k,l} \right) dx. \qquad (17)$$

We employ integration by parts to yield

$$\int_V \left( \frac{h_{ijkl}}{2} P_{i,j} u_{k,l} - \frac{h_{ijkl}}{2} P_i u_{k,jl} \right) dx + \text{Boundary terms}. \qquad (18)$$

Thus an energy density of the following form can be recovered

$$\frac{h_{ijkl}}{2} \left( P_{i,j} u_{k,l} - P_i u_{k,jl} \right) \qquad (19)$$

$\frac{\mathbf{h}}{2}$ can be redefined as $\mathbf{g}$ to recover the form of expression (12). Thus, instead of introducing two separate tensors $\mathbf{d}$ and $\mathbf{f}$, the enthalpy function can also be written as (Eliseev et al., 2009):

$$\Sigma = \frac{1}{2} a_{kl} P_k P_l + \frac{1}{2} b_{ijkl} P_{i,j} P_{k,l} + \frac{1}{2} c_{ijkl} e_{ij} e_{kl} + \frac{1}{2} h_{ijkl} \left( P_{i,j} u_{k,l} - P_k u_{i,jl} \right) \\ - \frac{1}{2} \varepsilon_0 \varphi_{,i} \varphi_{,j} + P_i \varphi_{,i}, \qquad (20)$$

where components of tensor $\mathbf{h}$ are combination of components of tensor $\mathbf{d}$ and tensor $\mathbf{f}$ which occur in the energy density described by Equation (8).

Standard variational analysis may now be employed to obtain a system of equilibrium equations, boundary conditions and constitutive relations for an isotropic material occupying domain $\Omega$ and bounded by a surface $S$. We omit these details as such deductions are routine. The major variables i.e. the electromechanical "stresses" are defined through the following relations:

$$\sigma_{ij} \equiv \frac{\partial \Sigma}{\partial e_{ij}}, \quad t_{ijm} \equiv \frac{\partial \Sigma}{\partial u_{i,jm}}$$
$$\Lambda_{ij} \equiv \frac{\partial \Sigma}{\partial P_{i,j}}, \quad \eta_i \equiv \frac{\partial \Sigma}{\partial P_i} \quad (21)$$

The definition of $\sigma_{ij}$ is the same as that of the stress tensor in classical elasticity; $\eta_i$ is the effective local electric force. The terms $t_{ijm}$ and $\Lambda_{ij}$ can be thought of as higher order stresses (moment stress) and higher order local electric force respectively. We now proceed to list the balance laws, boundary conditions and the constitutive relations.

(i) The Balance Laws:

$$\begin{aligned}
(\sigma_{ij} - t_{ijm,m})_{,j} + F_i &= 0 \\
-\Lambda_{ij,j} - \eta_i + \phi_{,i} &= 0 \\
-\varepsilon_0 \phi_{,ii} + P_{i,i} &= 0 \quad \text{in } \Omega \\
\phi_{,ii} &= 0 \quad \text{in } \Omega^*
\end{aligned} \quad \text{(22a-d)}$$

In Equations (22a-d), **F** is external body force. In the absence of the higher order stress $t_{ijm}$ which includes higher order gradients of the displacement vector (like $u_{i,jm}$), Equation (22a) reduces to the standard force balance equation of classical elasticity.

Since the term $\sigma_{ij} - t_{ijm,m}$ occurs in a force balance relation as evident in Equation (22a), we may interpret it as a "physical stress":

$$\sigma_{ij}^{phys} = \sigma_{ij} - t_{ijm,m} \quad (23)$$

(ii) The Boundary Conditions:

For all $x \in S$, the following conditions hold:

$$\begin{aligned}
n_i \sigma_{ij} &= T_j, \quad n_i \Lambda_{ij} = 0 \\
n_i \left( [\![\varepsilon_0 \phi_{,i}]\!] + P_i \right) &= 0 \\
[\![P_i]\!] &= 0
\end{aligned} \quad \text{(24a-d)}$$

**n** and **T** are the exterior normal unit vector and the surface traction vector respectively; $\varepsilon_0$ is the dielectric constant and the symbol $[\![\ ]\!]$ denotes the jump across the surface *S*. Equation (24d), i.e. continuity of polarization, is an extra condition that must be imposed to obtained a closed set of equations.

(iii) The Constitutive Relations:

$$\sigma_{ij} = c_{ijkl}e_{kl} + d_{klij}P_{k,l}$$
$$t_{ijm,m} = f_{kijm}P_{k,m}$$
$$\Lambda_{ij} = b_{ijkl}P_{k,l} + d_{ijkl}e_{k,l}$$
$$-E_i = a_{ij}P_j + f_{ijkl}u_{j,kl}$$

(25a-d)

Substituting the constitutive relations (25a-d) into the balance laws (22a-d), yields the Navier-like equations for dielectrics that incorporates the strain-polarization gradient coupling (Mindlin's theory) and the strain gradient – polarization coupling (flexoelectricity):

$$c_{44}\nabla^2\mathbf{u} + (c_{12} + c_{44})\nabla\nabla.\mathbf{u} + h_{12}\nabla^2\mathbf{P} + (h_{12} + h_{44})\nabla\nabla.\mathbf{P} + \mathbf{F} = 0,$$

$$h_{12}\nabla^2\mathbf{u} + (h_{12} + h_{44})\nabla\nabla.\mathbf{u} + (b_{44} + b_{77})\nabla^2\mathbf{P} +$$
$$(b_{12} + b_{44} - b_{77})\nabla\nabla.\mathbf{P} - a\mathbf{P} - \nabla\phi + \mathbf{E}^0 = 0,$$  (26a-c)

$$-\varepsilon_0\nabla^2\phi + \nabla.\mathbf{P} = 0.$$

## 3. Single Thin Film and Symmetry Arguments

Topologies of only certain symmetries can realize the central concept discussed in this paper. For example, isotropic spherical particles distributed in a matrix will not yield apparently piezoelectric composites even though the flexoelectric effect will cause local polarization fields. Due to spherical symmetry, the overall average polarization is zero. A similar composite but containing triangular shaped particles (and aligned in the same direction) will exhibit the required apparent piezoelectricity. Fabrication of the latter however is non-trivial.

In this section we explore symmetry considerations for the relatively easily manufacturable thin film based structures. Consider first a film made up of centrosymmetric material (Figure 1). More complex thin-film configurations solutions can be built using the elementary solution to be presented.

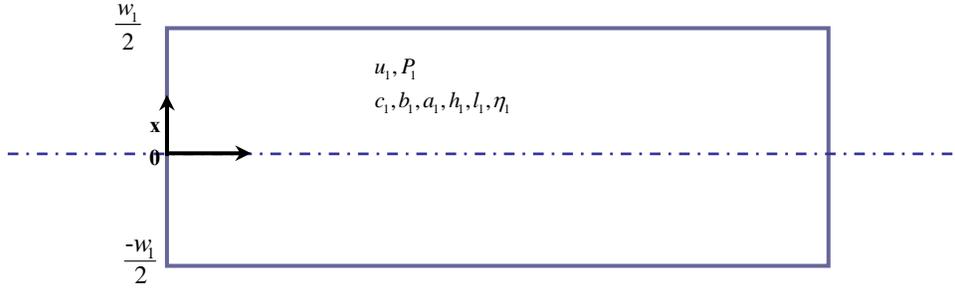

**Figure 1:** A thin film of a non-piezoelectric material e.g. paraelectric phase. BaTiO$_3$

For this film (idealized as a 1-D structure) the fields vary only in the x direction and the governing equations given by Equations (26a-c) simplify to

$$c\frac{\partial^2 u}{\partial x^2} + (d-f)\frac{\partial^2 P}{\partial x^2} = 0,$$

$$(d-f)\frac{\partial^2 u}{\partial x^2} + b\frac{\partial^2 P}{\partial x^2} - aP - \frac{\partial \phi}{\partial x} = 0, \qquad (27)$$

$$-\varepsilon_0 \frac{\partial^2 \phi}{\partial x^2} + \frac{\partial P}{\partial x} = 0.$$

Under open-circuit conditions, the electric displacement is zero.

$$-\varepsilon_0 \frac{\partial \phi}{\partial x} + P = 0. \qquad (28)$$

We arrive at the following equations:

$$\frac{bc - (d-f)^2}{c}\frac{\partial^2 P}{\partial x^2} - \left(a + \varepsilon_0^{-1}\right)P = 0$$

$$\Rightarrow \frac{\partial^2 P}{\partial x^2} - \frac{P}{l^2} = 0. \qquad (29\text{a-b})$$

where

$$l^2 = \frac{\left(bc - (d-f)^2\right)\varepsilon_0}{c\eta} \quad \text{and} \quad \eta = (1 + a\varepsilon_0). \qquad (30)$$

Equation (30b) can be solved for polarization to yield the form:

$$P = A_1 e^{(-\frac{x}{l})} + A_2 e^{(\frac{x}{l})}, \qquad (31)$$

where $A_1$ and $A_2$ are the constants of integration. The displacement field is:

$$u = A_3 + A_4 x + \frac{(d-f)}{c} e^{(-\frac{x}{l})} \left( A_1 + A_2 e^{(\frac{2x}{l})} \right). \tag{32}$$

Notice that in compliance with the Lifshitz invariance, the coefficients $d$ and $f$ appear together. For conciseness in the following sections, we write $h$ instead of $(d-f)$.

We also define the stress and the electric tensors respectively as:

$$\sigma = c\partial_x u + h\partial_x P,$$
$$\Lambda^{(ij)} = h\partial_x u + b\partial_x P. \tag{33a-b}$$

For the thin film in Figure (1), the following boundary conditions must be satisfied:

1. Applied stress boundary conditions

$$\left( c_1 \partial_x u_1 + h_1 \partial_x P_1 \right) = \sigma. \tag{34}$$

2. Electric tensor is set to zero at the free boundaries

$$\left( h_1 \partial_x u_1 + b_1 \partial_x P_1 \right) \Big|_{x \to \frac{w_1}{2}} = 0,$$
$$\left( h_1 \partial_x u_1 + b_1 \partial_x P_1 \right) \Big|_{x \to \frac{-w_1}{2}} = 0. \tag{35a-b}$$

3. Displacement u is set to zero at the origin

$$\left( u_1 \right) \Big|_{x \to 0} = 0. \tag{36}$$

Solving Equations (34-36) along with Equations (31-33), we obtain the expressions for polarization and displacement as:

$$P_1 = -\frac{\sigma \operatorname{sech}\left(\frac{w_1}{2l_1}\right) \sinh\left(\frac{x_1}{l_1}\right) h_1 \varepsilon_0}{c_1 l_1 \eta_1}$$

$$u_1 = \frac{\sigma}{c_1} \left( x + \frac{\sigma \operatorname{sech}\left(\frac{w_1}{2l_1}\right) \sinh\left(\frac{x_1}{l_1}\right) h_1^2 \varepsilon_0}{c_1 l_1 \eta_1} \right) \tag{37a-b}$$

The average polarization, as evident, is zero.

$$\frac{1}{w_1} \int_{-w_1/2}^{w_1/2} P_1 \, dx = 0 \tag{38}$$

To provide some physical perspective, we plot the polarization field for a 10 nm paraelectric $BaTiO_3$ (Figure 2). The applied stress is unity and the material constants are presented in Table 1 of Appendix I.

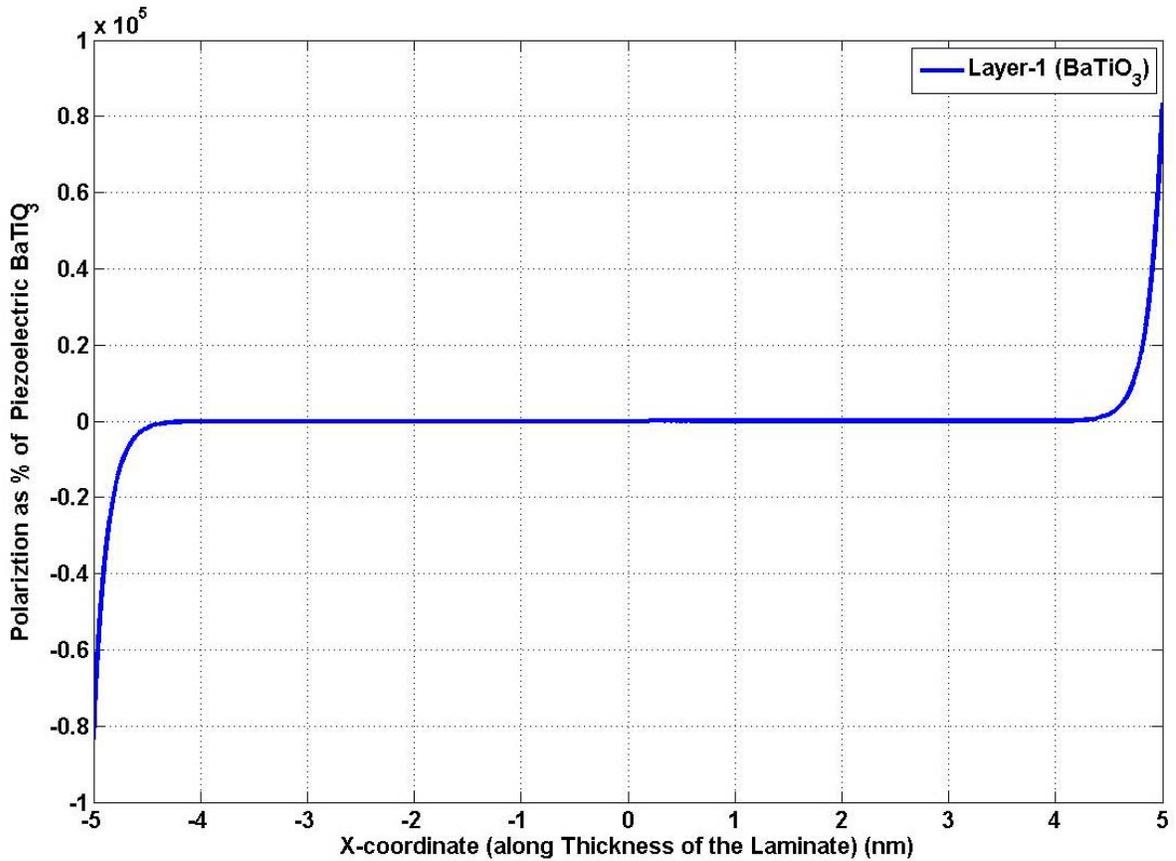

**Figure 3:** Polarization distribution in a thin film of paraelectric $BaTiO_3$
Total average polarization in the film is zero.

Building on the general solutions for the 1D mono-layer structure derived here, we can analyze various superlattices for the induced average polarization. Explicit expressions for induced polarization in each layer of the superlattice can be derived and used to calculate the averaged polarization in the entire composite.

A single thin film discussed so far is centrosymmetric about the mid-line. While a *finite* bilayer is non-centrosymmetric, a periodic two layered superlattice (a sequence of A-B-A-B-A-B….) is centrosymmetric. However, a tri-layer sequence e.g. A-B-C-A-B-C is non-centrosymmetric. In general, any odd-order stacking (of which A-B-C stacking is the simplest example) should yield a net non-zero average polarization.

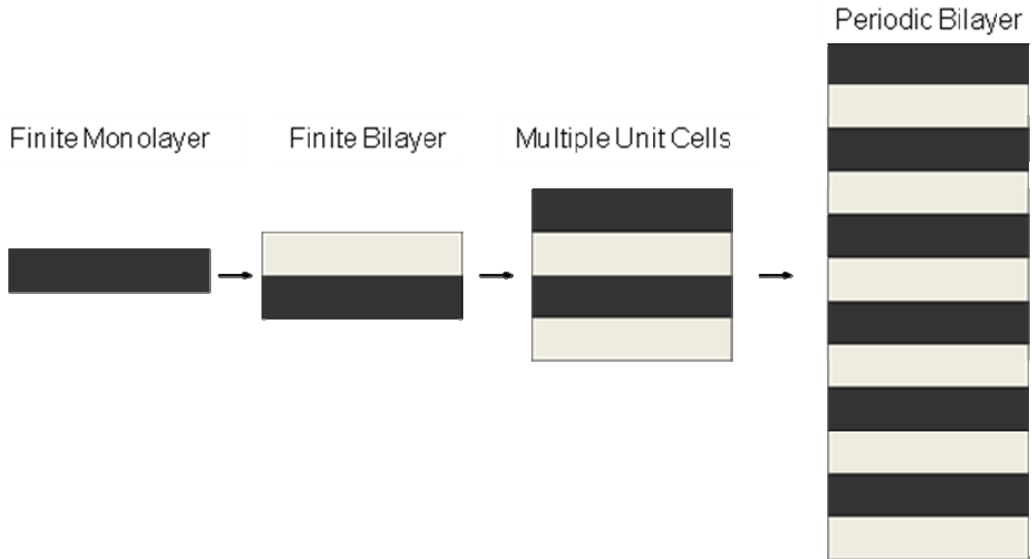

**Figure 3**: Apparently Piezoelectric Monolayer, Finite Bilayer and Periodic Bilayer. In case of periodic bilayer, the average polarization is zero.

Consider a periodic bilayer as shown in Figure 3. Each layer of a periodic bilayer experiences the strain gradients of same magnitudes in opposite directions at each interface. As a result of this 'inversion symmetry' of strain gradient the dipole moment induced in one layer of a unit cell is negated by the dipole moment induced in the next layer, rendering the overall average polarization in the composite to be zero. In other words the induced dipole moment in a layer negates the dipole moment induced in the adjacent layer. Thus overall average polarization in a periodic two layered superlattice is zero.

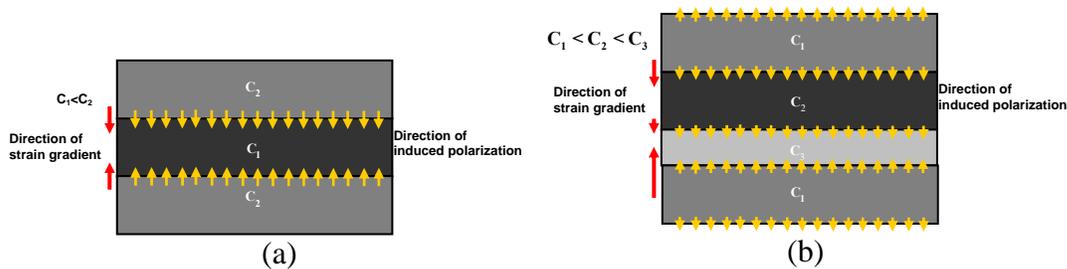

**Figure 4:** (a) Schematic of a periodic bilayer superlattice where induced dipole moment in a layer negates the dipole moment induced in the adjacent layer. Thus overall average polarization in a periodic bilayer superlattice is zero. (b) Schematic of a periodic trilayer superlattice shows that careful choice of material properties and superlatice topology can break the geometric centrosymmetry. Averaged strain gradients and thus the averaged induced polarization over the unit cell of a periodic trilayer superlattice are non-zero.

We must break this symmetry in order to get an apparent piezoelectric behavior in the periodic superlattices. The careful choice of material properties and superlattice topology can break the geometric centrosymmetry. If one introduces a third layer as shown in Figure 4 (b), the inversion symmetry is broken in such a periodic system. This periodic tri-layered superlattice thus is capable of generating a non-zero averaged polarization in the system.

## 4. Multilayer Thin Films and Superlattices

In an attempt to break the inherent centrosymmetry associated with a single thin film, we first consider a finite (non-periodic) bilyaer with thicknesses $w_1$ and $w_2$ as shown in the Figure 5:

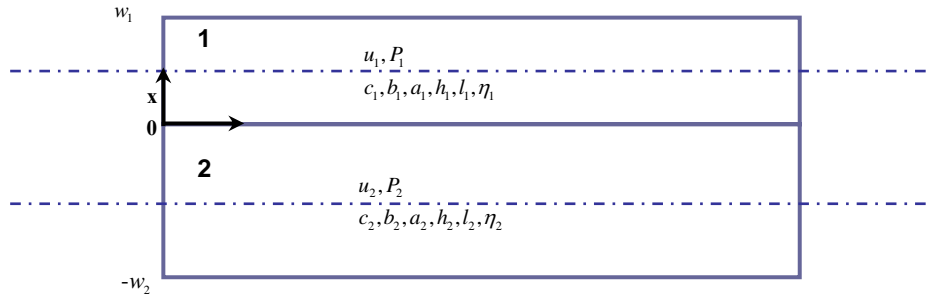

**Figure 5:** Unit cell of a bilayer film.

Even under application of uniform stress, change in material properties at the interface of the two layers will result in the presence of strain gradients in the system, which will induce polarization due to the flexoelectric coupling. Note that in the finite case, such structures will in fact lack the inversion symmetry of individual layers around the interface of the two layers. Thus, we would expect a nonzero average polarization under suitable boundary conditions. As derived in the previous section, the polarization and displacement in layer-1 is of the form

$$P_1 = A_{11} \exp\left(-\frac{x}{l_1}\right) + A_{12} \exp\left(\frac{x}{l_1}\right)$$

$$u_1 = A_{13} + A_{14}x + \frac{h_1}{c_1}\exp\left(-\frac{x}{l_1}\right)\left(A_{11} + A_{12}\exp\left(2\frac{x}{l_1}\right)\right).$$

(39a-b)

Similarly, in layer-2, the polarization and displacement is given by

$$P_2 = A_{21} \exp\left(-\frac{x}{l_2}\right) + A_{22} \exp\left(\frac{x}{l_2}\right)$$

$$u_2 = A_{23} + A_{24}x + \frac{h_2}{c_2}\exp\left(-\frac{x}{l_2}\right)\left(A_{21} + A_{22}\exp\left(2\frac{x}{l_2}\right)\right).$$

(40a-b)

The following boundary conditions must be satisfied:
1. Applied stress boundary conditions

$$\left(c_1 \partial_x u_1 + h_1 \partial_x P_1\right) = \sigma,$$
$$\left(c_2 \partial_x u_2 + h_2 \partial_x P_2\right) = \sigma.$$

(41a-b)

2. Continuity of stress at the interface

$$[\![\sigma]\!]_{x\to 0} = \left(\sigma^{(1)}\big|_{x\to 0} - \sigma^{(2)}\big|_{x\to 0}\right) = 0.$$

(42)

This condition is redundant, since in this case, the previous two (applied stress) conditions trivially ensure this continuity.

3. Displacements at the interface are zero.

$$u_1\big|_{x\to 0} = 0,$$
$$u_2\big|_{x\to 0} = 0.$$

(43a-b)

4. Electric tensor $E_{ij}$ is set to zero at the free boundaries

$$\Lambda_{ij}^{(1)} = \left(h_1 \partial_x u_1 + b_1 \partial_x P_1\right)\big|_{x\to w_1} = 0,$$
$$\Lambda_{ij}^{(2)} = \left(h_2 \partial_x u_2 + b_2 \partial_x P_2\right)\big|_{x\to -w_2} = 0.$$

(44a-b)

5. The Electric tensor ($E_{ij}$) is specified to be continuous (but not necessarily zero) at the interface

$$[\![\Lambda_{ij}]\!]_{x\to 0} = \left(\Lambda^{(1)}_{ij}\big|_{x\to 0} - \Lambda^{(2)}_{ij}\big|_{x\to 0}\right) = 0.$$

(45)

6. Polarization (P) is specified to be continuous at the interface

$$[\![P]\!]_{x\to 0} = \left(P_1\big|_{x\to 0} - P_2\big|_{x\to 0}\right) = 0.$$

(46)

Unlike classical theory of piezoelectricity, an additional boundary condition is required at the interface on the polarization field.

We finally obtain the following results:

$$P_1 = -\frac{\sigma\epsilon_0 \left( \begin{array}{c} -\cosh\left(\frac{x-w_1}{l_1}\right)\left(-1+\cosh\left(\frac{w_2}{l_2}\right)\right)c_1 h_2 l_1 \eta_1 \\ +c_2 h_1 \left( \begin{array}{c} \left(-\cosh\left(\frac{x}{l_1}\right)+\cosh\left(\frac{x-w_1}{l_1}\right)\right)\cosh\left(\frac{w_2}{l_2}\right)l_1 \eta_1 \\ -\sinh\left(\frac{x}{l_1}\right)\sinh\left(\frac{w_2}{l_2}\right)l_2 \eta_2 \end{array} \right) \end{array} \right)}{c_1 c_2 l_1 \eta_1 \left( \cosh\left(\frac{w_2}{l_2}\right)\sinh\left(\frac{w_1}{l_1}\right)l_1 \eta_1 + \cosh\left(\frac{w_1}{l_1}\right)\sinh\left(\frac{w_2}{l_2}\right)l_2 \eta_2 \right)}.$$

(47)

and

$$P_2 = \frac{\sigma\epsilon_0 \left( \begin{array}{c} \left(-1+\cosh\left(\frac{w_1}{l_1}\right)\right)\cosh\left(\frac{x+w_2}{l_2}\right)c_2 h_1 l_2 \eta_2 \\ +c_1 h_2 \left( \begin{array}{c} -\sinh\left(\frac{w_1}{l_1}\right)\sinh\left(\frac{x}{l_2}\right)l_1 \eta_1 \\ +\cosh\left(\frac{w_1}{l_1}\right)\left(\cosh\left(\frac{x}{l_2}\right)-\cosh\left(\frac{x+w_2}{l_2}\right)\right)l_2 \eta_2 \end{array} \right) \end{array} \right)}{c_1 c_2 l_2 \eta_2 \left( \cosh\left(\frac{w_2}{l_2}\right)\sinh\left(\frac{w_1}{l_1}\right)l_1 \eta_1 + \cosh\left(\frac{w_1}{l_1}\right)\sinh\left(\frac{w_2}{l_2}\right)l_2 \eta_2 \right)}.$$

(48)

Average polarization in the superlattice is calculated to be:

$$\frac{1}{w_1+w_2}\left( \int_0^{w_1} P_1 dx + \int_{-w_2}^{0} P_2 dx \right) =$$

$$\frac{4\sigma \sinh\left(\frac{w_1}{2l_1}\right)\sinh\left(\frac{w_2}{2l_2}\right)\epsilon_0 (\eta_1-\eta_2) \left( \begin{array}{c} \cosh\left(\frac{w_1}{2l_1}\right)\sinh\left(\frac{w_2}{2l_2}\right)c_1 h_2 l_1 \eta_1 \\ +\cosh\left(\frac{w_2}{2l_2}\right)\sinh\left(\frac{w_1}{2l_1}\right)c_2 h_1 l_2 \eta_2 \end{array} \right)}{c_1 c_2 (w_1+w_2)\eta_1 \eta_2 \left( \cosh\left(\frac{w_2}{l_2}\right)\sinh\left(\frac{w_1}{l_1}\right)l_1 \eta_1 + \cosh\left(\frac{w_1}{l_1}\right)\sinh\left(\frac{w_2}{l_2}\right)l_2 \eta_2 \right)}.$$

(49)

We note here that the average polarization directly depends on the difference between the dielectric constants of the constituent materials. Larger differences between the dielectric constants of the two layers leads to a larger induced average polarization, which translates into a stronger apparent piezoelectric behavior. Numerical results for BaTiO$_3$-MgO bilayer are shown in Figure (6). For these results, we assume both layers, layer-2 (MgO) and layer-1 (BaTiO$_3$) to be 10nm thick subject to a unit applied stress.

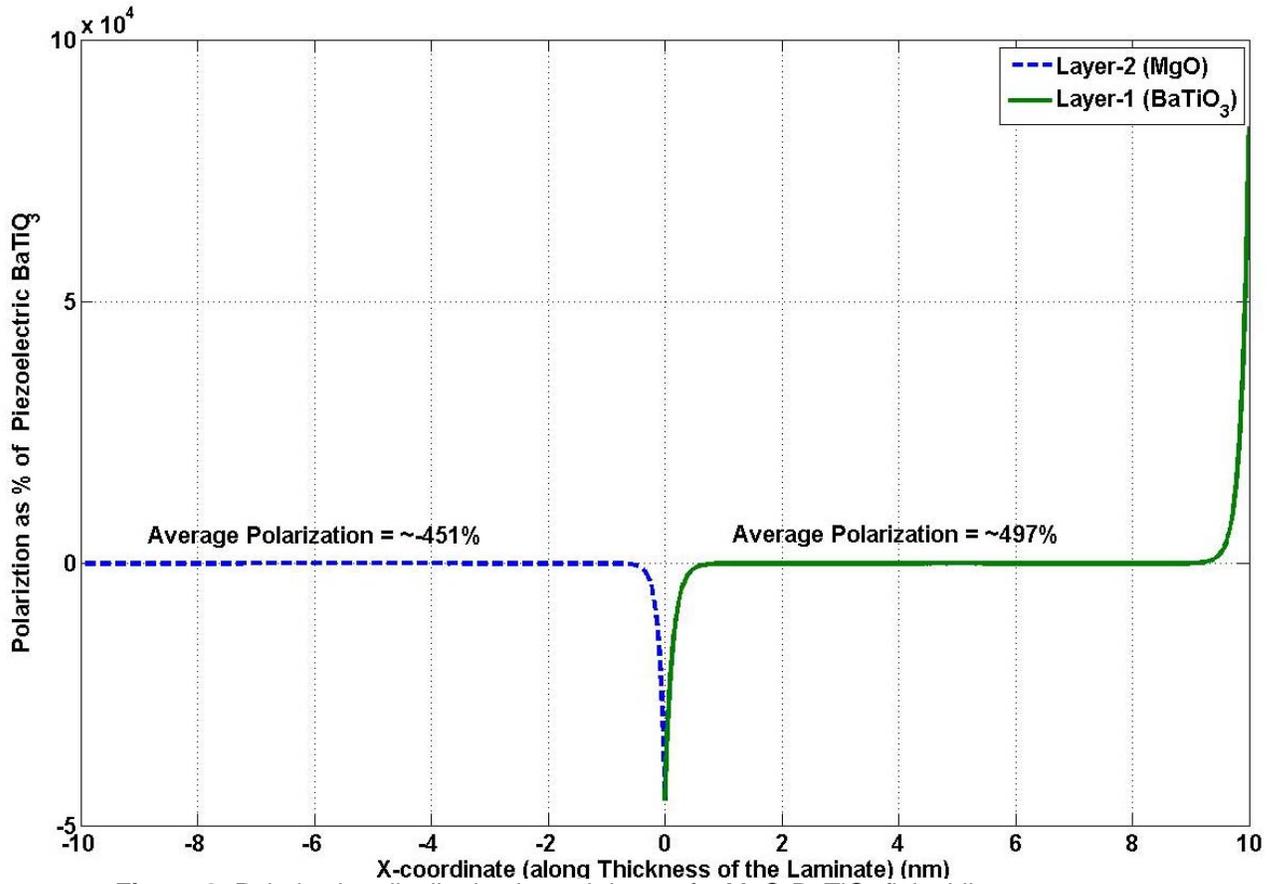

**Figure 6:** Polarization distribution in each layer of a MgO-BaTiO$_3$ finite bilayer. Total average polarization in the bilayer is 23% of piezoelectric BaTiO$_3$.

In the case of a finite two layered film structure made up of non-piezoelectric materials the averaged net polarization is nonzero and for the numerical results shown in Figure (6) we obtain an effective piezoelectric constant of about 23 % of BaTiO$_3$----a well known piezoelectric material.

### 4.1 Periodic Two Layered Superlattices

Consider a *periodic* bilayer superlattice (A-B-A-B sequence). In addition to the boundary conditions presented earlier, we impose periodicity requirement:

$$P_1\Big|_{x \to w_1} = P_2\Big|_{x \to -w_2},$$
$$E_{ij}^{(1)}\Big|_{x \to w_1} = E_{ij}^{(2)}\Big|_{x \to -w_2}. \tag{50a-b}$$

The final results are:

$$P_1 = \frac{\sigma \sinh\left(\frac{-2x+w_1}{2l_1}\right)\sinh\left(\frac{w_2}{2l_2}\right)(c_2 h_1 - c_1 h_2)\epsilon_0}{c_1 c_2 \left(\cosh\left(\frac{w_1}{2l_1}\right)\sinh\left(\frac{w_2}{2l_2}\right)l_1 \eta_1 + \cosh\left(\frac{w_2}{2l_2}\right)\sinh\left(\frac{w_1}{2l_1}\right)l_2 \eta_2\right)}. \tag{51}$$

and

$$P_2 = \frac{\sigma \sinh\left(\frac{w_1}{2l_1}\right)\sinh\left(\frac{2x+w_2}{2l_2}\right)(c_2 h_1 - c_1 h_2)\epsilon_0}{c_1 c_2 \left(\cosh\left(\frac{w_1}{2l_1}\right)\sinh\left(\frac{w_2}{2l_2}\right)l_1 \eta_1 + \cosh\left(\frac{w_2}{2l_2}\right)\sinh\left(\frac{w_1}{2l_1}\right)l_2 \eta_2\right)}. \tag{52}$$

As expected from symmetry arguments, the overall average polarization is zero (see Figure 7).

$$\frac{1}{w_1 + w_2}\left(\int_0^{w_1} P_1 dx + \int_{-w_2}^{0} P_2 dx\right) = 0. \tag{53}$$

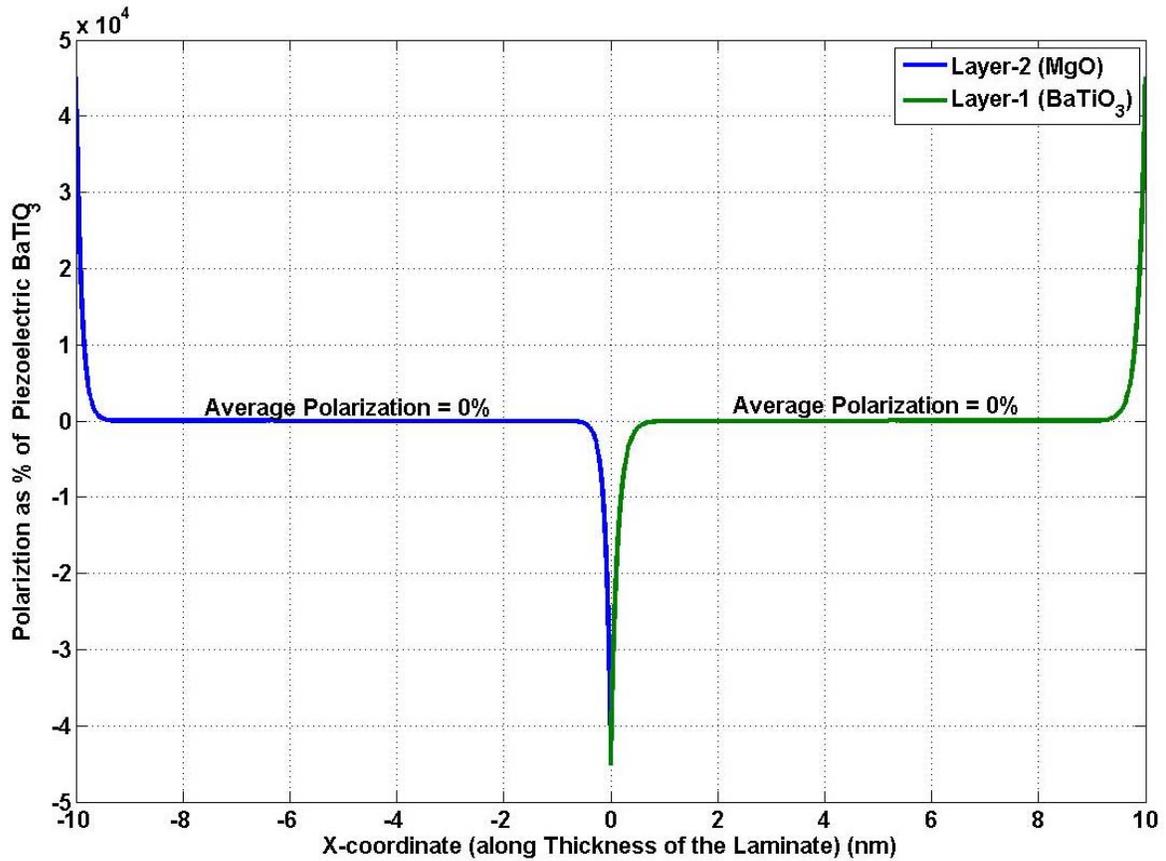

**Figure 7:** Polarization in each layer of a MgO-BaTiO$_3$ Periodic Two Layered Superlattice.

## 4.2 Periodic Trilayer

Consider a periodic trilayer as shown in the Figure 8.

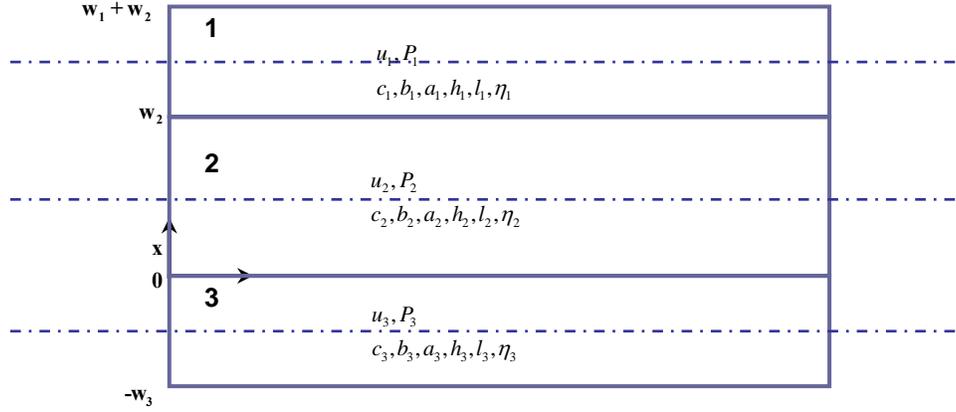

**Figure 8:** Unit cell of a Three Layered Superlattice

Note that the origin is defined at interface of layer-2 and layer-3. The general forms of the polarization and displacement fields are:

$$P_1 = A_{11} \exp\left(-\frac{x}{l_1}\right) + A_{12} \exp\left(\frac{x}{l_1}\right),$$

$$u_1 = A_{13} + A_{14}x + \frac{h_1}{c_1} \exp\left(-\frac{x}{l_1}\right)\left(A_{11} + A_{12} \exp\left(2\frac{x}{l_1}\right)\right),$$

$$P_2 = A_{21} \exp\left(-\frac{x}{l_2}\right) + A_{22} \exp\left(\frac{x}{l_2}\right),$$

$$u_2 = A_{23} + A_{24}x + \frac{h_2}{c_2} \exp\left(-\frac{x}{l_2}\right)\left(A_{21} + A_{22} \exp\left(2\frac{x}{l_2}\right)\right), \quad (54\text{a-f})$$

$$P_3 = A_{31} \exp\left(-\frac{x}{l_3}\right) + A_{32} \exp\left(\frac{x}{l_3}\right),$$

$$u_3 = A_{33} + A_{34}x + \frac{h_3}{c_3} \exp\left(-\frac{x}{l_3}\right)\left(A_{31} + A_{32} \exp\left(2\frac{x}{l_3}\right)\right).$$

Boundary conditions are essentially the same as in the previous section. The expressions for polarization in each layer in this case are rather complex (although closed-form), hence only numerical results are presented here. We consider a 'SrTiO$_3$-MgO-BaTiO$_3$' three layered superlattice. We take each layer to be 10 nm thick and magnitude of the applied stress to be unity.

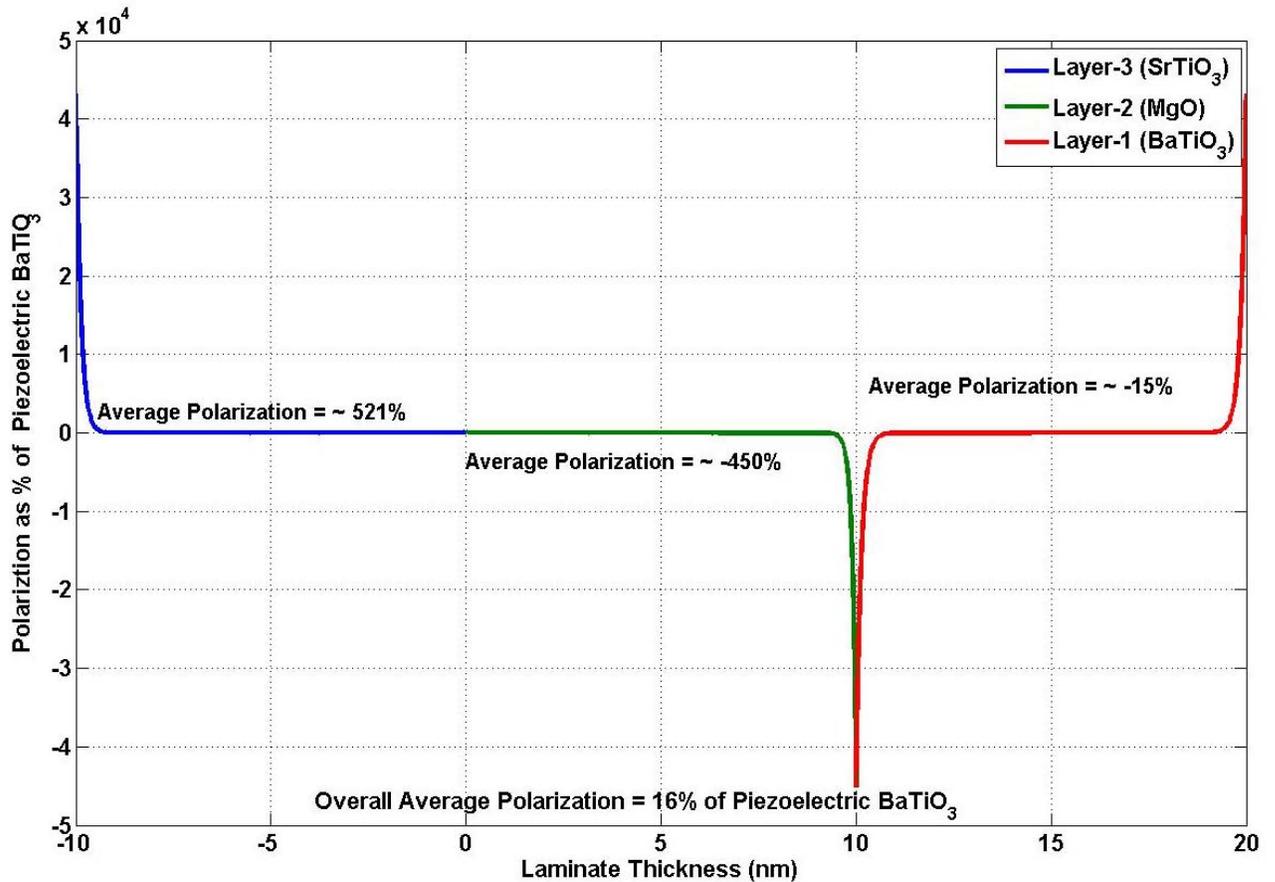

**Figure 9:** Polarization in each layer of a SrTiO$_3$-MgO-BaTiO$_3$ Periodic Three Layered Superlattice. Overall average polarization in the superlattice is 16% of the piezoelectric BaTiO$_3$.

## 4.3 Effect of layer sizes on Induced Average Polarization

Induced average polarization in the periodic three layered superlattice can be fine tuned by controlling the sizes of each layer. We wish to maximize the average polarization in the superlattice with respect to the size of each layer, such that total thickness of the superlattice unit cell does not exceed 20 nm. We restrict the minimum size of each layer to 2 nm. These size-restrictions are based on limitations imposed by current capabilities of state-of-art fabrication processes of ceramic materials. Figure 10 depicts the polarization profiles for various layer thicknesses.

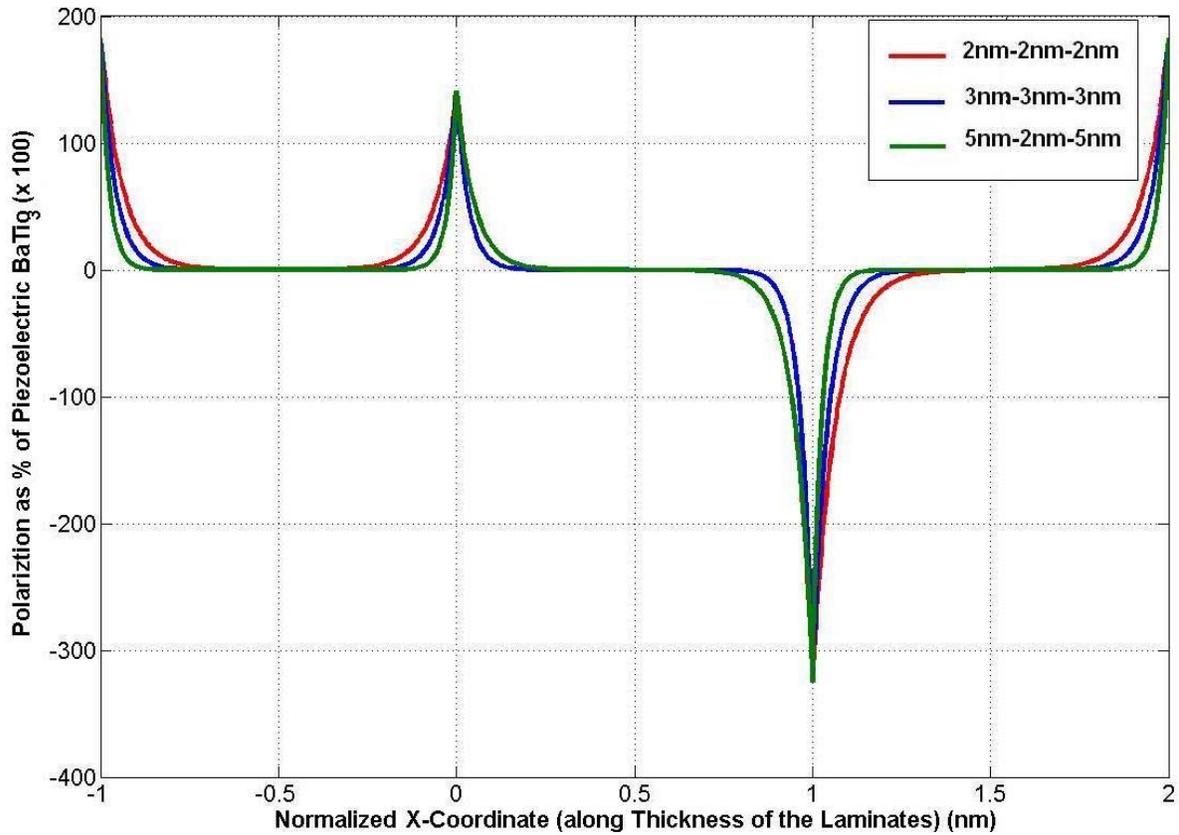

**Figure 10:** Polarization in each layer of a $SrTiO_3$-$MgO$-$BaTiO_3$ Periodic Three Layered Superlattice for various layer thicknesses.

Since the average polarization is inversely proportional to the layer size, we expect that the average polarization will be maximum for smallest possible layer thickness. The solution to this problem in fact confirms this expectation and we obtain a maximum average polarization of 77.5% of piezoelectric $BaTiO_3$ when each layer is 2 nm thick.

## 5. Concluding Remarks

In this paper, we provide exact results for flexoelectric response of thin films under stress and structures based on thin films such as periodic super-lattices. The interplay between thin film thickness, symmetry (represented in this context by stacking sequence), and flexoelectricity allows the tantalizing possibility of creating manufacturable apparently piezoelectric thin film super-lattices without using piezoelectric materials. In one scenario (trilayer sequence of BTO and MgO with thicknesses in the range of 2 nm), close of 75 % of the value of ferroelectric BTO is obtainable.

Acknowledgements: P.S and N.S acknowledge financial support from the NSF under NIRT Grant Nos. CMMI 0708096 and IMI center IIMEC--DMR 0844082 (www.iimec.tamu.edu). C.M.L. gratefully acknowledges support from ONR through contract N00014-07-1-0469

# Appendix: Material Properties

Maranganti and Sharma (2009) recently calculated flexoelectric properties were atomistically for several dielectrics which agree with the experimental estimates (Ma and Cross, 2001b; 2002; 2003, 2006; Fu et al 2006, Zubko, 2007) to an order of magnitude. However, for the case of BaTiO$_3$, a large discrepancy with the experimental estimates was observed, reasons for which are still not fully understood. It should be noted that in this current work, we have used the experimental estimates for calculations.

|  |  | BaTiO3 | MgO | SrTiO3 |
|---|---|---|---|---|
| $p_{33}$ | $(C/N)$ | $7.80 \times 10^{-11}$ | – | $3.00 \times 10^{-14}$ |
| Lattice Parameter ($a$) | $(A°)$ | 4.00 | 4.21 | 3.91 |
| Relative Permittivity | $(-)$ | $4.00 \times 10^3$ | 9.70 | $3.00 \times 10^2$ |
| $b_{11}$ | $(Nm^4/C^2)$ | $6.77 \times 10^{-06}$ | $5.67 \times 10^{-08}$ | $4.14 \times 10^{-06}$ |
| $c_{11}$ | $(N/m^2)$ | $1.62 \times 10^{11}$ | $3.00 \times 10^{11}$ | $3.50 \times 10^{11}$ |
| $h_{11}$ | $(Nm/C)$ | $-1.55 \times 10^{05}$ | $1.29 \times 10^{02}$ | $-1.20 \times 10^{03}$ |
| $l_1$ | $(A°)$ | 1.30 | 1.00 | 1.20 |